# Double Minimum Variance Beamforming Method to Enhance Photoacoustic Imaging

ROYA PARIDAR[1], MOEIN MOZAFFARZADEH[1,2], MOHAMMADREZA NASIRIAVANAKI[3], AND MAHDI OROOJI[1*]

[1]*Department of Biomedical Engineering, Tarbiat Modares University, Tehran, Iran.*
[2]*Research Center for Biomedical Technologies and Robotics (RCBTR), Institute for Advanced Medical Technologies (IAMT), Tehran, Iran.*
[3]*Department of Biomedical Engineering, Wayne State University, Detroit, MI, USA.*
[*]*Corresponding author: morooji@modares.ac.ir*

One of the common algorithms used to reconstruct photoacoustic (PA) images is the non-adaptive Delay-and-Sum (DAS) beamformer. However, the quality of the reconstructed PA images obtained by DAS is not satisfying due to its high level of sidelobes and wide mainlobe. In contrast, adaptive beamformers, such as minimum variance (MV), result in an improved image compared to DAS. In this paper, a novel beamforming method, called Double MV (D-MV) is proposed to enhance the image quality compared to the MV. It is shown that there is a summation procedure between the weighted subarrays in the output of the MV beamformer. This summation can be interpreted as the non-adaptive DAS beamformer. It is proposed to replace the existing DAS with the MV algorithm to reduce the contribution of the off-axis signals caused by the DAS beamformer between the weighted subarrays. The numerical results show that the proposed technique improves the full-width-half-maximum ($FWHM$) and signal-to-noise ratio ($SNR$) for about 28.83 $\mu m$ and 4.8 $dB$ in average, respectively, compared to MV beamformer. Also, quantitative evaluation of the experimental results indicates that the proposed D-MV leads to 0.15 $mm$ and 1.96 $dB$ improvement in $FWHM$ and $SNR$, in comparison with MV beamformer.    © 2018 Optical Society of America

*OCIS codes:*    (110.5120) Photoacoustic imaging; (110.3000) Image quality assessment; (110.2990) Image formation theory; (110.3010) Image reconstruction techniques.

## 1. INTRODUCTION

Photoacoustic imaging (PAI) is a non-invasive medical imaging modality which is fast growing [1]. It combines the physics of the ultrasound (US) and the optical imaging modalities, and provides the resolution of US and contrast of the optical imaging [2, 3]. In PAI, a non-ionizing radiations is used in order to illuminate the medium and generate the acoustic waves. Compared to the optical imaging, the advantage of PAI is its high penetration depth. In addition, it does not have the speckle noise appeared in US imaging [4]. PAI provides functional information since the optical absorption depends on the physiological conditions, such as hemoglobin concentration [5, 6]. PAI has several applications such as ocular imaging [7], structural and functional imaging [8], whole body of small animal imaging [9] and blood flow measurement [10]. In PAI, the imaging method is divided into two categories; Photoacoustic Microscopy (PAM) and Photoacoustic Tomography (PAT) [11]. In PAT, which is suitable for *in vivo* imaging, an array of ultrasonic transducers is used to receive the generated acoustic signals propagated from the medium [12]. Then, a reconstruction algorithm is used in order to reconstruct the absorption distribution in the medium [13]. Recently, low-cost PAT and PAM systems are extensively investigated [14–18]. Due to the high similarity between the US and PAI, beamforming algorithms used in US imaging can be applied on the PAI, with some modifications [19]. One of the most common reconstruction algorithm used in US and PAI, is Delay-and-Sum (DAS). However, the quality of the reconstructed Photoacoustic (PA) images are not satisfying, having a wide mainlobe and high level of sidelobes. Delay-Multiply-and-Sum (DMAS), introduced by Matrone *et al.* , was proposed in order to improve the quality of the reconstructed images compared to DAS [20]. Some algorithms have been developed to improve the contrast of the DMAS algorithm [21–23]. Adaptive beamformers, which are commonly used in Radar, weigh the detected signals based on the available information in the signals. Therefore the adaptive beamformers would improve the image quality compared to non-adaptive beamformers. One of the adaptive beamforming methods that has mostly been developed in medical applications, is minimum variance (MV) algorithm [24], in which the weights are adopted to the inverse covariance matrix of the received data. It has been shown that in MV beamformer, the large sidelobes are steered in the directions where the received energy is low, and zeros are placed in directions where there is interference [25]. Using this algorithm, the quality of the reconstructed image will be significantly improved in terms of mainlobe width and sidelobe levels compared to non-adaptive DAS beamformer. Significant improvements in the image quality have been seen by applying some modifications to the MV beamformer. In [26], temporal averaging has been employed to improve the image contrast. Another novel beamforming algorithm, named MV-Based DMAS (MVB-DMAS) was proposed for linear-array PAI in which the MV beamformer is used inside the expansion of the non-adaptive DMAS beamformer [27, 28]. It has been shown that the quality of the reconstructed PA images are improved compared to MV and DMAS. Coherence factor (CF) has been applied to the MV beamformed signals to improve the resolution and suppress the sidelobes, and it has been shown that this technique is suitable in high frame rate (HFR) imaging applications [29]. Two modifications of CF have been introduced for PAI in order to have a lower sidelobes and higher resolution, compared to the conventional CF [30, 31]. Eigenspace-Based Minimum Variance (EIBMV) and forward-backward (FB) MV beamformers also have been applied to medical US imaging to improve the image quality and robustness [32, 33]. EIBMV was combined with DMAS to further improve the PA image quality [34, 35].

In this paper, a novel technique, named Double-MV (D-MV) is introduced using the same concept introduced in [21, 28]. It can be seen that there is a summation procedure between the weighted subarrays in the output of the MV beamformer, which is similar to the non-adaptive DAS beamformer. We propose to apply the MV to the weighted subarrays instead of the existing DAS beamformer to enhance the quality of the reconstructed PA images. It is shown that the proposed technique results in a narrower mainlobe and lower level of sidelobes compared to MV.

The rest of the paper is organized as follows. In section 2, a brief explanation about the concepts of the PAI and MV beamforming method along with the proposed algorithm are presented. The numerical and experimental results are presented in sections 3 and 4, respectively. Finally, the discussion and the conclusion are reported in the section 5 and section 6, respectively.

## 2. METHODS AND MATERIALS

### A. Fundamental of Photoacoustics

In PAI, a short laser pulse is used to irradiate the medium. The light is absorbed proportional to the characteristic of the medium. Following the light absorption, the thermoelastic expansion is occurred and results in the PA wave propagation with a frequency range of hundreds of megahertz. Then, the propagated PA signals are received by an ultrasonic transducer. The key problem is to reconstruct the absorption distribution of the medium from the received PA signals, using an appropriate algorithm. There are several methods and algorithms for PA image reconstruction, especially in the case of linear-array transducer [21, 27, 28, 34]. It should be noted that there are some similarities between US and PA in terms of acoustic wave measurement at the surface of the medium [36]. Therefore, the PA images can be reconstructed using the algorithms and beamformers used in US image formation. In the following, MV beamformer is explained. More information about the fundamental of PA and wave equation is reported in [37].

### B. Minimum Variance Beamformer

Consider a linear-array transducer consisting of $M$ elements. The output of the MV beamformed signal, $y(k)$, is written as below:

$$y(k) = \sum_{m=1}^{M-1} w_m(k) x_m(k - \Delta_m(k)), \quad (1)$$

where $x_m(k)$ is the received signal (delayed proportional to the distance of the element $m$ to the point target), $\Delta_m$ and $k$ are the time delay for $m^{th}$ detector and the time index, respectively, and $w_m(k)$ is the calculated weight corresponding to the delayed received signals. As it can be seen in (1), MV is considered as the sum of the weighted delayed signals. The weight is calculated in a way that the signal-to-interference-plus-noise ratio ($SINR$) would be maximized [38]:

$$\text{SINR} = \frac{\sigma_s^2 |W(k)^H a|^2}{W(k)^H R(k) W(k)}, \quad (2)$$

where $\sigma_s^2$ is the signal power and $R(k)$ is the spatial covariance matrix of the interference-plus-noise. To maximize the $SINR$, the denominator of (2) should to be minimized while the unit gain is retained at the focal point:

$$\min_{W(k)} \quad W(k)^H R(k) W(k) \text{ s.t. } W(k)^H a = 1, \quad (3)$$

where $W(k) = [w_1(k), w_2(k), \cdots, w_m(k)]^T$, $a$ is the steering vector and $(.)^H$ denotes the conjugate transpose operator. It should be noted that since the received signals are delayed, $a$ is considered as a vector of ones. Equation (3) has the following solution using Lagrange multiplier method:

$$W(k) = \frac{R(k)^{-1} a}{a^H R(k)^{-1} a}. \quad (4)$$

Referring to [25], MV steers the large sidelobes in directions where the received energy is low, and places zero in directions where there is an interference. It is not possible to obtain the exact spatial covariance matrix in practical applications since the signals in the medical US and PAI cases are non-stationary, and therefore the spatial covariance matrix should be estimated using $N$ received samples:

$$\hat{R} = \frac{1}{N} \sum_{k=1}^{N} \bar{X}(k) \bar{X}(k)^H, \quad (5)$$

where $\bar{X}(k) = [x_1(k), x_2(k), \cdots, x_M(k)]^T$ is the array of the delayed received signals. A technique, named spatial smoothing, is used to obtain a good estimation of the spatial covariance matrix; the array sensor is divided into some overlapping subarrays with the length of $L$. Then, $R(k)$ is estimated by averaging the calculated covariance matrices of each subarray. The estimated covariance matrix is written as below:

$$\hat{R}(k) = \frac{1}{(M-L+1)} \sum_{l=1}^{M-L+1} \bar{X}_l(k) \bar{X}_l(k)^H, \quad (6)$$

where $\bar{X}_l(k) = [x_l(k), x_{l+1}(k), \cdots, x_{l+L-1}(k)]^T$ is the delayed received signals of the $l^{th}$ subarray. In other words, the received signals obtained from $M - L + 1$ subarrays are replaced by $N$ received samples in (5). It should be noted that the smaller $L$ results in a more robust estimation, but the resolution would be decreased. Therefore, the length of the subarray determines the trade off between the resolution and the robustness. Also, a more robust estimation would be achieved by applying diagonal loading (*DL*) in which $\Delta.trace\{R\}$ is added to the estimated covariance matrix, before the weights are calculated. $\Delta$ is a constant parameter that corresponds to the subarray length and is much smaller than $\frac{1}{L}$ [39]. In addition, the temporal averaging over $2K + 1$ samples can be used in spatial covariance matrix estimation in order to achieve a better resolution while the contrast is retained [26]:

$$\hat{R}(k) = \frac{1}{(2K+1)(M-L+1)} \times \sum_{n=-K}^{K} \sum_{l=1}^{M-L+1} \bar{X}_l(k+n) \bar{X}_l(k+n)^H. \quad (7)$$

Using $\hat{R}(k)$ instead of $R(k)$ in (4), the weight is estimated, and the output is obtained from the following equation:

$$\tilde{y}(k) = \frac{1}{M-L+1} \sum_{l=1}^{M-L+1} W(k)^H \bar{X}_l(k). \quad (8)$$

### C. Proposed Method

In MV beamformer, the weighted subarrrays are summed to make the output beamformed signals. This summation procedure is similar to the DAS beamformer. In this paper, it is proposed to apply MV to the weighted subarrays instead of the existing DAS algebra. To better illustration, consider the expansion of (8) as below:

$$\tilde{y}(k) = \frac{1}{M-L+1} \sum_{l=1}^{M-L+1} W(k)^H \bar{X}_l(k) = \frac{W(k)}{M-L+1} \left[ X_1(k) + X_2(k) + \cdots + X_{M-L+1}(k) \right]. \quad (9)$$

Equation (9) can be rewritten as follows:

$$\tilde{y}(k) = p_1(k) + p_2(k) + \cdots + p_{M-L+1}(k)$$
$$= \sum_{i=1}^{M-L+1} p_i(k), \quad (10)$$

where

$$p_i(k) = \frac{1}{M-L+1} W(k)^H X_i(k). \quad (11)$$

From the new form of the MV output, shown in (10), the possibility of applying MV to the weighted subarrays is more clearly identified. From (10), the existing DAS algebra between the weighted subarrays, $p_i(k)$, is obtained. As mentioned before, DAS is a non-adaptive beamforming method, and it cannot suppress the off-axis signals and noise as well as the adaptive beamformers. Therefore, the quality of the reconstructed image obtained from this algorithm is not satisfying, having wide mainlobe and high level of sidelobes. It is proposed to apply MV instead of the existing DAS between the weighted subarrays inside the MV formula to generate the new D-MV technique. It is expected that the quality of the reconstructed PA image be improved by replacing MV with DAS algorithm. To apply D-MV, the new spatial covariance matrix and consequently the new weight should to be estimated between the weighted subarrays, named $\hat{R}_D(k)$ and $W_D(k)$, respectively; Each weighted subarray, $p_i(k)$, is considered as an element, and therefore, it is possible to divide them into some new overlapping subarrays, with the length of $L_D$. Then, the new spatial covariance matrix for the weighted subarrays would be estimated by the following equation:

$$\hat{R}_D(k) = \frac{1}{(2K_D+1)(M_D-L_D+1)} \times \sum_{n=-K_D}^{K_D} \sum_{l=1}^{M_D-L_D+1} \bar{P}_l(k+n) \bar{P}_l(k+n)^H, \quad (12)$$

where $M_D = M - L + 1$ is the total number of subarrays, which are considered as elements, and $\bar{P}_l(k) = \left[ p_l(k), p_{l+1}(k), \cdots, p_{l+L_p-1}(k) \right]^T$. Temporal averaging over $2K_D + 1$ samples also can be applied in this stage. The final estimated output beamformed signal, $\tilde{y}_D(k)$, is obtained from the following equation:

$$\tilde{y}_D(k) = \frac{1}{M_D-L_D+1} \sum_{l=1}^{M_D-L_D+1} W_D(k)^H \bar{P}_l(k), \quad (13)$$

where the new weight is estimated by replacing $\hat{R}_D(k)$ with $R(k)$ in (4). It is expected to achieve an enhanced reconstructed PA image using this new beamforming method. The numerical and experimental results, presented in the next section, show that the D-MV beamformer improves the reconstructed images in terms of mainlobe width and sidelobe levels, compared to MV.

## 3. NUMERICAL RESULTS

### A. Imaging Setup

The k-wave MATLAB toolbox is used to design the absorbers and the linear array sensor [40]. An imaging region is simulated with 70 *mm* in the vertical axis and 20 *mm* in the lateral axis. Nine spherical absorbers with a radius of 0.1 *mm* are located as the initial pressures. The absorbers are centered on the lateral axis and located along the vertical axis. There is 5 *mm* vertical distance between each two absorbers, and the positioning of the point targets are started from the distance of 25 *mm* from the array sensor. The speed of sound is assumed 1540 *m/s*. A linear-array consisting of 128 elements is used to receive the propagated PA signals from the point targets. The central frequency of the array is 5 *MHZ* with 77% bandwidth. The subarray lengths are considered $L = M/2$ and $L_D = (M-L)/2$. *DL* is applied to the spatial covariance matrixes in each stage of applying D-MV with the assumption of $\Delta = 1/100L$ and $\Delta_D = 1/100L_D$ in the first and the second stage, respectively. Temporal averaging is applied over 7 received samples($K = 3$) in the first stage, and no temporal averaging is applied in the second stage of applying D-MV ($K_D = 0$). Also, Gaussian noise is added to the received signals to make the signals similar to the practical condition, having a signal-to-noise Ratio (*SNR*) of 50 *dB*. The Hilbert transform is performed to the signals received by the array elements. Then, the normalization and log-compression procedure are performed after the reconstruction algorithm is applied.

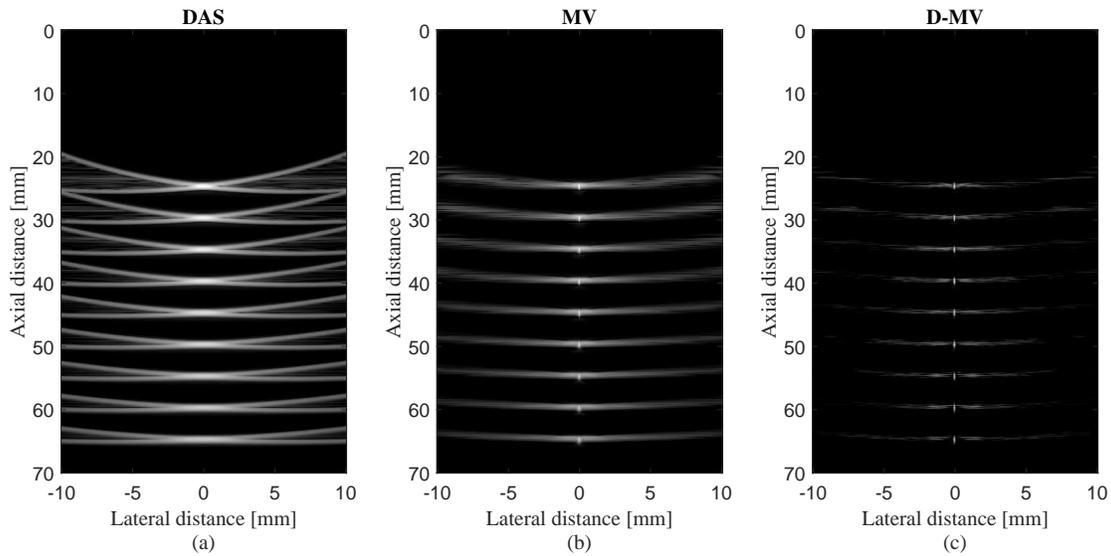

**Fig. 1.** The reconstructed PA images of 9 point targets using (a) DAS, (b) MV, (c) D-MV. Noise is added to the received signals having a $SNR$ of 50 $dB$. $K = 3, K_D = 0$.

## B. Qualitative Evaluation

Fig. 1 shows the reconstructed PA images of the point targets. The reconstructed images using DAS, MV and D-MV are depicted in Fig. 1(a), Fig. 1(b) and Fig. 1(c), respectively. It is obvious from the images that the adaptive MV beamformer outperforms the non-adaptive DAS beamformer, and the quality of the reconstructed image is improved. Comparing the proposed D-MV algorithm with the MV, it can be concluded that D-MV results in an improved reconstructed image in the terms of width of mainlobe and noise suppression. To have a better evaluation, the lateral variations are presented in Fig. 2. The lateral variations at the depth of 25 $mm$, 45 $mm$, 50 $mm$ and 65 $mm$ are shown in Fig. 2(a), Fig. 2(b), Fig. 2(c) and Fig. 2(d), respectively. As demonstrated by the lateral variations, the both adaptive beamformers outperform DAS in terms of mainlobe width and sidelobes. Also, the proposed D-MV beamformer would reconstruct the PA images with a significantly narrower width of mainlobe and lower level of sidelobes, compared to the MV beamformer and DAS.

## C. Quantitative Evaluation

$SNR$ and the spatial resolution are two important metrics to evaluate and compare the performance of the beamformers. The full-width-half-maximum ($FWHM$) can be used to estimate the spatial resolution. The calculated $FWHMs$ in each depth of positioning point targets (shown in Fig. 1) are presented in Table 1. The calculated $FWHMs$ show that the largest $FWHM$ value belongs to DAS beamformer, at the all depths, as expected. Comparing the $FWHM$ values of point targets achieved by MV and the proposed D-MV beamformers, it can be concluded that D-MV results in a narrower width of mainlobe in -6 $dB$, which means that the D-MV beamformer would provide a reconstructed image with a better resolution compared to MV beamformer, as it can be seen from the lateral variations in Fig. 2. It is worth to mention that the difference between the calculated $FWHMs$ at different depths of positioning points, is almost negligible using D-MV, in comparison with other beamformers; The

**Table 1.** The calculated $FWHMs$ ($\mu m$) at the different depths of Fig. 1.

| Depth ($mm$) | DAS | MV | Double MV |
|---|---|---|---|
| 25 | 1144.1 | 118.1 | 100.3 |
| 30 | 1416.7 | 119.2 | 100.4 |
| 35 | 1719.3 | 120 | 100.6 |
| 40 | 1968.2 | 125 | 100.7 |
| 45 | 2288.5 | 130.7 | 101 |
| 50 | 2657.8 | 132.4 | 101.3 |
| 55 | 2999.9 | 133.1 | 101.4 |
| 60 | 3431.9 | 136.9 | 101.7 |
| 65 | 3679.6 | 153.3 | 101.8 |

difference between the calculated $FWHMs$ at the first and the last depth of positioning points, is 2.53 $mm$, 35.2 $\mu m$ and 1.5 $\mu m$ for DAS, MV and D-MV, respectively. As a result, the resolution remains almost constant in all depths of imaging using the proposed D-MV, compared to DAS and MV beamformers. $SNR$ is calculated as follows:

$$SNR = 20\log_{10} P_{signal}/P_{noise}, \qquad (14)$$

where $P_{signal}$ is the difference between the maximum and the minimum intensity of the reconstructed image, and $P_{noise}$ is the standard deviation of the reconstructed image [21, 23]. The $SNR$ in each depth of positioning point targets are calculated and presented in Table 2. It can be seen from the calculated $SNRs$ that DAS beamformer results in lower $SNR$ compared to MV beamformer; existence of high sidelobes in the reconstructed image obtained by DAS results in a lower $SNR$ compared to MV. Since the proposed D-MV beamformer performs the MV twice, it leads to a better noise reduction, and therefore, higher $SNR$,

">Research Article | Journal of the Optical Society of America A | 5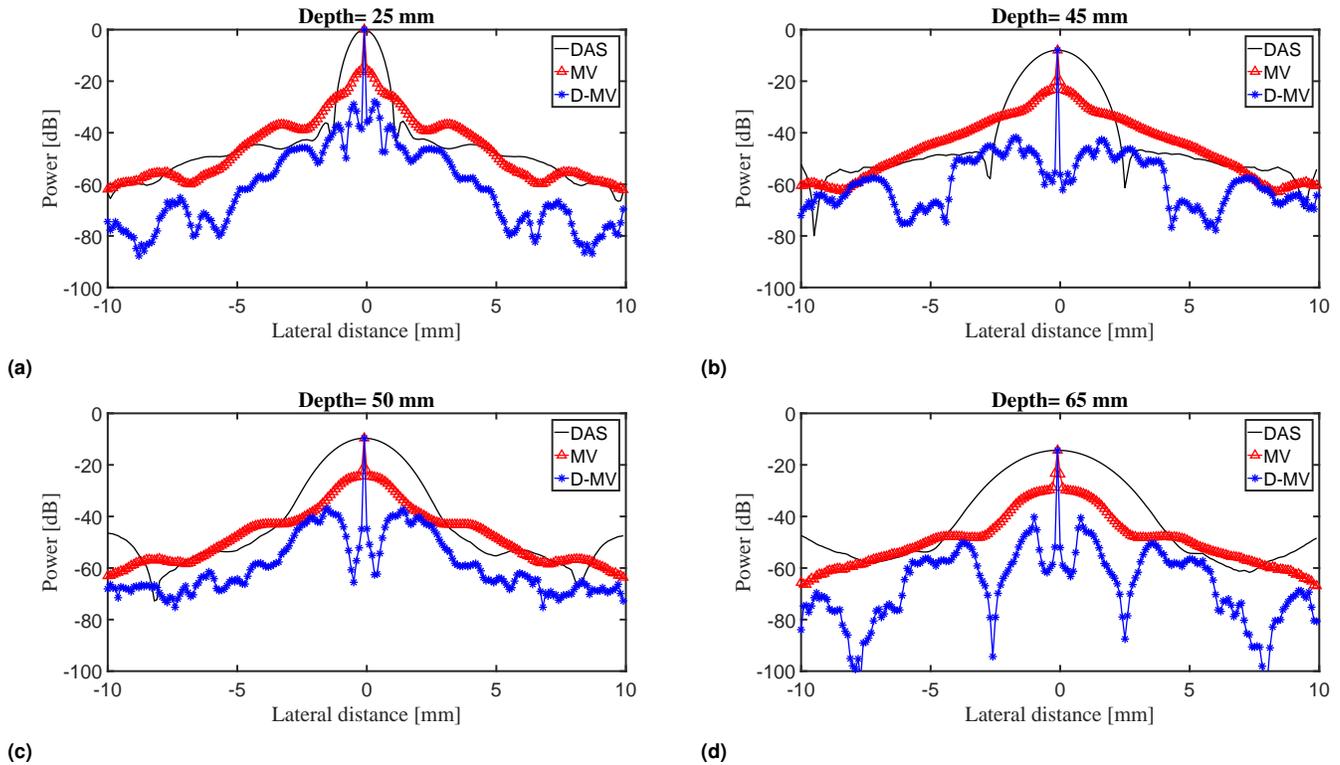

**Fig. 2.** The lateral variations of the images shown in Fig. 1 at the depth of (a) 25 *mm*, (b) 45 *mm*, (c) 50 *mm* and (d) 65 *mm*. Noise is added to the received signals having a $SNR$ of 50 *dB*.

**Table 2.** The calculated $SNR$s (*dB*) for the Fig. 1.

| Depth (*mm*) | DAS | MV | D-MV |
|---|---|---|---|
| 25 | 73.57 | 81.15 | 89.26 |
| 30 | 71.80 | 78.83 | 84.70 |
| 35 | 69.29 | 76.57 | 79.46 |
| 40 | 67.23 | 73.74 | 77.01 |
| 45 | 66.45 | 68.67 | 71.58 |
| 50 | 65.11 | 66.24 | 71.51 |
| 55 | 63.62 | 64.76 | 69.73 |
| 60 | 61.94 | 62.51 | 69.28 |
| 65 | 56.66 | 62.33 | 65.55 |

compared to MV.

### D. Imaging at the Presence of High level of Noise

To evaluate the performances of the beamformers in terms of noise reduction, the reconstruction was performed while a high level of Gaussian noise was added to the detected PA waves, resulting in a $SNR$ of 10 *dB*. The reconstructed images of Nine point targets simulation are shown in Fig. 3, at the presence of a high level of noise. The reconstructed images using DAS, MV and the proposed D-MV algorithms are shown in Fig. 3(a), Fig. 3(b) and Fig. 3(c), respectively. As shown, the quality of the reconstructed image obtained from DAS is more affected by the presence of noise. Also, D-MV results in an improved reconstructed image in terms of noise reduction, compared to DAS and MV. Fig 4(a) and Fig. 4(b) show the lateral variations of the point targets in which 10 *dB* noise is added to the received signals, at the depth of 30 *mm* and 55 *mm*, respectively, to compare the sidelobes. It can be seen that the mainlobe width and the sidelobe levels of the reconstructed image obtained from D-MV are enhanced, compared to other beamformers, at the presence of high level of noise.

## 4. EXPERIMENTAL RESULTS

### A. Experimental Setup

A system consisting of an ultrasound data acquisition system, Vantage 128 Verasonics (Verasonics, Inc., Redmond, WA) and a Q- switched Nd:YAG laser (EverGreen Laser, Double-pulse Nd: YAG system) is used for PAI. The pulse repetition rate of the laser is 25 *Hz* with 532 *nm* wavelength and 10 *ns* pulse width. A linear array sensor (L7-4, Philips Healthcare) consisting of 128 elements with 5.2 *MHZ* central frequency is used to receive the propagated PA waves from the designed phantom. A function generator is used to synchronize all operations (i.e., laser firings and PA signal recording). The data sampling rate is 20.8320 *MHZ*. The schematic of the designed PAI system is presented in Fig. 5. It should be mentioned that the transducer was perpendicular to the wires. Thus, it is expected to see a cross section of the wires which would look like a point target. In the following, the reconstructed images of the designed wire phantom are presented.

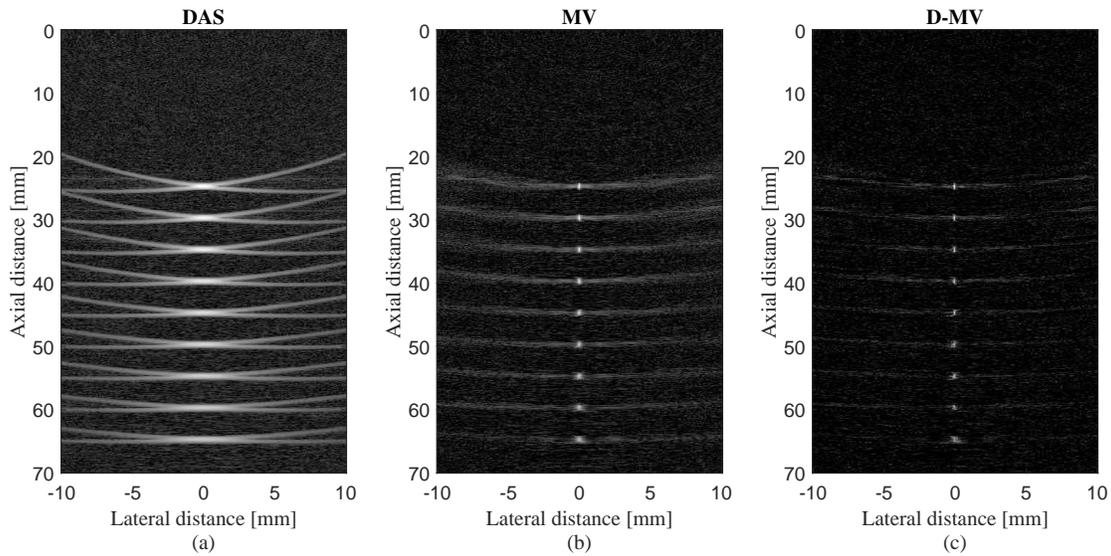

**Fig. 3.** The reconstructed PA images of 9 point targets using (a) DAS, (b) MV, (c) D-MV. Noise is added to the received signals having a *SNR* of 10 *dB*. $K, K_D = 0$

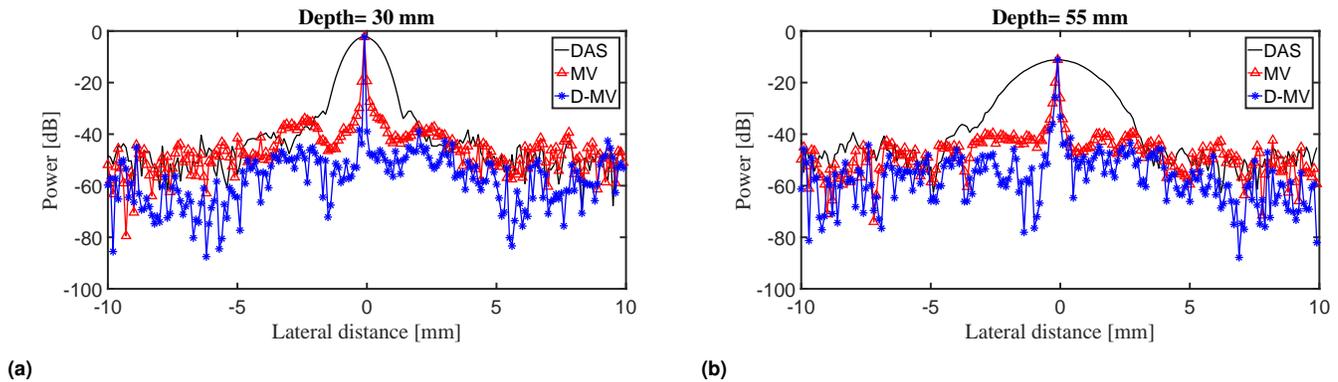

**Fig. 4.** The lateral variations of the images shown in Fig. 3 at the depth of (a) 30 *mm* and (b) 55 *mm*. Noise is added to the received signals having a *SNR* of 10 *dB*.

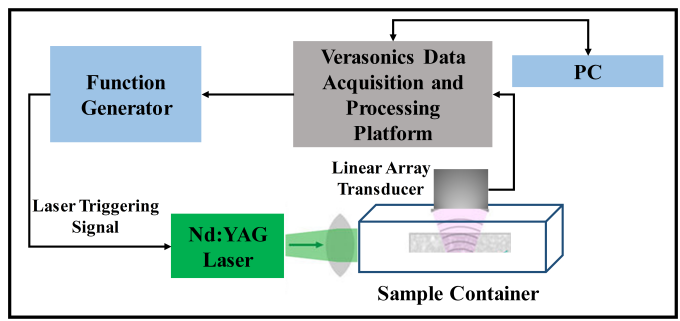

**Fig. 5.** Schematic of the experimental setup.

### A.1. Wire phantom

The reconstructed images of the designed wire phantom experiments are presented in Fig. 6 and Fig. 7. The reconstructed images are shown with a dynamic range of 40 *dB*. It should be noted that the surface of the array sensor is perpendicular to the designed wire phantom during the imaging procedure, and therefore, the targets are reconstructed like a point. It can be seen in Fig. 6 and Fig. 7 that the reconstructed image using the non-adaptive DAS beamformer leads to larger width of mainlobe compared to MV beamformer. In other words, MV improves the resolution of the reconstructed image compared to DAS, as expected. The narrowest mainlobe width, and consequently, the highest resolution would be achieved by the proposed D-MV beamformer, shown in Fig. 6(c) and Fig. 7(c). To have a better evaluation, the lateral variations are presented in Fig. 8 and Fig. 9. As demonstrated using the circlesi and arrows, the sidelobes are further degraded using the proposed method. Also, the *SNR* and *FWHM* for the reconstructed images shown in Fig. 6 and Fig. 7 are calculated and presented in Table 3 and Table 4, respectively. It is obvious that D-MV beamformer improves the SNR since the sidelobes gained by this algorithm are reduced,

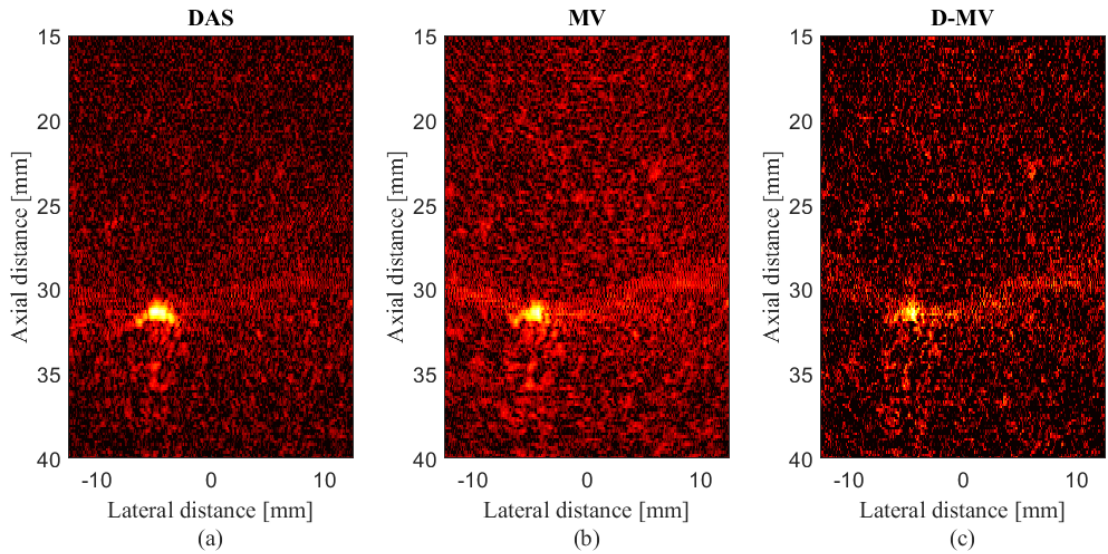

**Fig. 6.** The reconstructed experimental images using (a) DAS, (b) MV and (c) D-MV beamromers. The wire phantom was used at the target. All the images are shown with a dynamic range of 40 $dB$.

and it provides a higher noise suppression, in comparison with DAS and MV.

## 5. DISCUSSION

The major benefit of the proposed D-MV beamformer is the image enhancement in terms of mainlobe width and sidelobe levels, compared to MV beamformer. The reconstructed images obtained from DAS algorithm is not satisfying since this algorithm is non-adaptive and the weights are predefined (data-independent). MV, as an adaptive beamformer, calculates the weights based on the characteristics of the detected signals. Therefore, it results in the resolution improvement compared to DAS. From the expansion of the formula related to the output of the MV beamformed signals, (8), and rewriting it in a new form shown in (10), it can be seen that the DAS algebra exists between the weighted subarrays. By replacing MV with the existing DAS between the weighted subarrays, the quality of the MV beamformed signals would be significantly improved. In fact, the proposed D-MV algorithm re-weights the weighted delayed signals adaptively, and the resolution of the reconstructed PA images are significantly enhanced compared to MV. As it can be seen in Fig. 1(b) and Fig. 1(c), where the image reconstruction is performed by MV and D-MV, respectively, the proposed D-MV beamformer improved the image quality compared to MV. Also, more noise reduction is occurred. The lateral variations shown in Fig. 2 indicate that the width of mainlobe is significantly reduced using D-MV in all the depths of imaging. D-MV also results in the reconstructed image with a better quality at the presence of a high level of noise, as it can be seen from Fig. 3, where 10 $dB$ noise is added to the received signals. The reconstructed image using MV beamformer, has a lower quality due to the existence of the DAS algebra between the weighted subarrays and its high off-axis signals contribution. On the other hand, the proposed D-MV beamformer leads to a more noise reduction and a higher quality reconstructed image. The effects of the image enhancement using the proposed D-MV can be seen in the experimental results shown in Fig. 6 in which the resolution is improved compared to MV. It should be noted that the computa-tional complexity of the D-MV beamformer is higher compared to other beamformers. The computational complexity of DAS beamformer is $O(M)$. The processing time that MV needs to calculate the weights is $O(M^2)$. It is obvious from the output of the MV formula that the calculated weight is multiplied to the summation of the subarrays. Therefore, the total computational complexity of MV beamformer is $O(M^2M) = O(M^3)$. The proposed D-MV beamformer consists of two stages; in the first stage, the processing time for calculating the weights is $O(M^2)$. In the second stage, MV is applied to the weighted subarrays with the computational complexity of $O(M^3)$. Therefore, total computational complexity of D-MV would be $O(M^2M^3) = O(M^5)$, which is higher in comparison with MV beamformer. Of note, there are a large number of investigation to reduce the complexity of the MV beamformer. Having the computational burden of the MV beamformer reduced would result in imposing lower computational burden by the proposed algorithm.

**Table 3.** The calculated $SNR$s ($dB$) and $FWHM$ ($mm$) for experimental results shown in Fig 6.

| Beamformer | DAS | MV | D-MV |
|---|---|---|---|
| $FWHM$ | 0.72 | 0.66 | 0.51 |
| $SNR$ | 42.46 | 44.07 | 46.03 |

**Table 4.** The calculated $SNR$s ($dB$) and $FWHM$ ($mm$) for experimental results shown in Fig 7.

| Beamformer | DAS | MV | D-MV |
|---|---|---|---|
| $FWHM$ | 0.83 | 0.35 | 0.24 |
| $SNR$ | 14.57 | 21.74 | 27.87 |

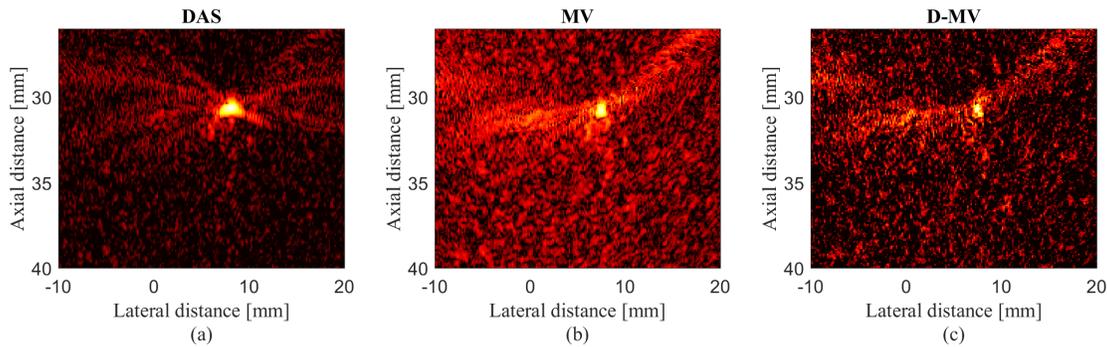

**Fig. 7.** The reconstructed experimental images using (a) DAS, (b) MV and (c) D-MV beamromers. The wire phantom was used at the target. All the images are shown with a dynamic range of 40 *dB*.

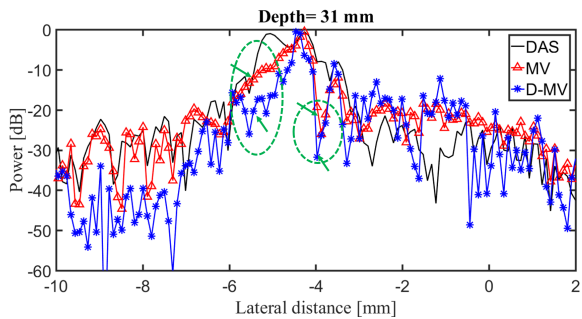

**Fig. 8.** The lateral variations of the images shown in Fig. 6 at the depth of 30 *mm*.

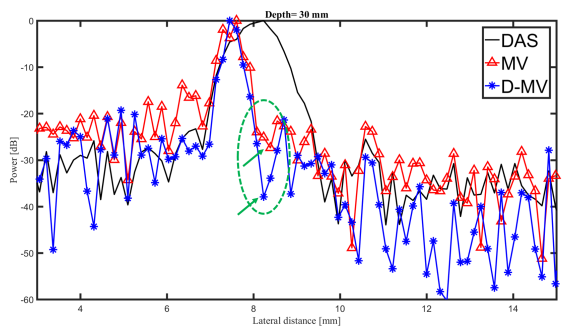

**Fig. 9.** The lateral variations of the images shown in Fig. 7 at the depth of 30 *mm*.

## 6. CONCLUSION

DAS is one of the commonly used algorithms in PA image reconstruction due to its simple implementation. However, the DAS beamformed signals result in a wide mainlobe and large sidelobes. In MV, as an adaptive beamformer, the calculated weights are changed depending on the characteristics of the signals. Therefore, the resolution of the reconstructed image would be significantly enhanced compared to DAS. By the expansion of the MV formula, it can be seen that DAS algebra exists between the weighted subarrays. In this work, a novel beamforming method, named D-MV, was proposed to reconstruct the PA images, where MV is replaced with the existing DAS algebra. To put it more simply, it was applied to the weighted subarrays in the output of the MV beamformed signals. The numerical and experimental results showed that the proposed D-MV technique improved the images in terms of mainlobe width and sidelobe levels compared to MV. The quantitative evaluation of the numerical results, achieved by D-MV, indicated that $FWHM$ and $SNR$ were improved for about 28.83 $\mu m$ and 4.8 $dB$, compared to MV, respectively.